%% file: Template_Regular.tex
\documentclass{article}
\usepackage{cite}
\usepackage{spconf,amsmath,graphicx,url}
\usepackage{color, colortbl}
\usepackage{xcolor}
\colorlet{mono}{blue!10}
\colorlet{regional}{teal!10}
\colorlet{multi}{yellow!10}
\usepackage{booktabs}
\usepackage{multicol, multirow}
\usepackage{url}
\usepackage{hyperref}
\usepackage{pgfplots}
\pgfplotsset{compat=1.16}

\definecolor{color1}{RGB}{255,0,0}     
\definecolor{color2}{RGB}{0,255,0}     
\definecolor{color3}{RGB}{0,0,255}     
\definecolor{color4}{RGB}{255,255,0}   
\definecolor{color5}{RGB}{255,0,255}   
\definecolor{color6}{RGB}{0,255,255}   
\definecolor{color7}{RGB}{192,192,192} 
\definecolor{color8}{RGB}{128,128,128} 
\definecolor{color9}{RGB}{128,0,0}     
\definecolor{color10}{RGB}{128,128,0}  
\definecolor{color11}{RGB}{0,128,0}    
\definecolor{color12}{RGB}{128,0,128}  
\definecolor{color13}{RGB}{0,128,128}  
\definecolor{color14}{RGB}{0,0,128}    
\definecolor{color15}{RGB}{139,69,19}  
\definecolor{color16}{RGB}{255,20,147} 
\definecolor{color17}{RGB}{255,165,0}  
\definecolor{color18}{RGB}{85,107,47}  
\definecolor{color19}{RGB}{70,130,180} 
\definecolor{color20}{RGB}{106,90,205} 

\title{Findings of the 2023 ML-SUPERB Challenge: Pre-Training and Evaluation over More Languages and Beyond}
\name{
    \begin{tabular}[c]{@{}c@{}c@{}c@{}c@{}}
    Jiatong Shi$^1$, William Chen$^1$, Dan Berrebbi$^1$, Hsiu-Hsuan Wang$^2$, Wei-Ping Huang$^2$, \\ En-Pei Hu$^2$, Ho-Lam Chuang$^2$, Xuankai Chang$^1$, Yuxun Tang$^3$, Shang-Wen Li$^4$,  \\ Abdelrahman Mohamed$^5$, Hung-yi Lee$^2$, Shinji Watanabe$^1$
    \end{tabular}
    }
\address{$^1$Carnegie Mellon University, $^2$National Taiwan University  \\$^3$Renmin University of China, $^4$Meta AI, $^5$Rembrand}
\copyrightnotice{979-8-3503-0689-7/23/\$31.00~\copyright2023 IEEE}

\begin{document}
\ninept
\maketitle

\begin{abstract}
The 2023 Multilingual Speech Universal Performance Benchmark (ML-SUPERB) Challenge expands upon the acclaimed SUPERB framework, emphasizing self-supervised models in multilingual speech recognition and language identification. The challenge comprises a research track focused on applying ML-SUPERB to specific multilingual subjects, a Challenge Track for model submissions, and a New Language Track where language resource researchers can contribute and evaluate their low-resource language data in the context of the latest progress in multilingual speech recognition. The challenge garnered 12 model submissions and 54 language corpora, resulting in a comprehensive benchmark encompassing 154 languages. The findings indicate that merely scaling models is not the definitive solution for multilingual speech tasks, and a variety of speech/voice types present significant challenges in multilingual speech processing.
\end{abstract}
\begin{keywords}
Multilingual speech recognition, self-supervised learning, ML-SUPERB
\end{keywords}
\section{Introduction}
\label{sec:intro}

Self-supervised learning (SSL) has gained significant popularity in the speech community due to its effectiveness in capturing essential speech features, such as phonemes and acoustic units, through training on large amounts of unlabeled speech data \cite{mohamed2022self}. These SSL models have shown remarkable improvements in various downstream tasks, including speech recognition, speaker identification, and emotion recognition \cite{yang21c_interspeech}. In recent years, researchers have proposed diverse SSL models with different training objectives, operating under various data conditions, model architectures, and modalities \cite{baevski2020wav2vec, hsu2021hubert}.

The Speech Universal PERformance Benchmark (SUPERB), established in 2021, has emerged as a popular benchmark for evaluating speech SSL representations \cite{yang21c_interspeech}. The primary goal of this benchmark is to compare speech SSLs across various speech processing tasks, encompassing aspects such as content, speaker, semantics, and paralinguistics. However, one limitation of SUPERB is its exclusive focus on English speech in its downstream tasks. In contrast, there is a growing interest in applying SSL models to multilingual scenarios, including training multilingual SSL models \cite{babu2021xls, conneau2020unsupervised, duquenne2022speechmatrix} or utilizing SSL models in a cross-lingual manner \cite{zhao2022improving, berrebbi22_interspeech, wu2020self,  li2022asr2k}. To facilitate research in these areas, a new benchmark called multilingual SUPERB (ML-SUPERB) has been proposed \cite{shi2023ml}.

ML-SUPERB has been designed to encompass a wide range of languages, including both high-resource languages like English and endangered languages such as Totonac or Mixtec \cite{shi2021highland, shi2021leveraging, amith_audio_corpus_sierra, amith_yoloxochitl_mixtec, amith_totonac}. The benchmark primarily focuses on evaluating SSL models for automatic speech recognition (ASR) and language identification (LID). To cater to different use cases for SSL models, ML-SUPERB includes two tracks with four different tasks: the monolingual track (monolingual ASR) and the multilingual track (multilingual ASR, LID, joint multilingual ASR/LID). Similar to SUPERB, ML-SUPERB utilizes frozen SSL models as feature extractors and employs a lightweight downstream model that can be fine-tuned for different tracks to achieve high training efficiency. The released public benchmark of ML-SUPERB covers 143 languages, making it highly inclusive and representative of diverse linguistic contexts.

Following the release of ML-SUPERB, the ML-SUPERB Challenge was inaugurated. In addition to a research track encompassing a variety of topics, we implemented two tracks for competitors: the Challenge Track and the freshly minted New Language Track. The Challenge Track emphasizes challenge performance, analogous to the SUPERB SLT2022 challenge. Conversely, the New Language Track instigates a novel design to the challenge by inviting participants to contribute their language resources. Through the creation of the New Language Track, ML-SUPERB continues to evolve by integrating new languages into its framework.

The challenge drew considerable interest and participation, with the Challenge Track receiving 12 model submissions and the New Language Track gaining an additional 54 valuable language resources. In incorporating these new languages, ML-SUPERB now extends its reach to an impressive total of \textbf{154} languages. It's noteworthy that all contributions were exclusively from academic institutions, demonstrating that the realm of multilingual SSL research isn't confined to large corporations, and indeed, academia can exert substantial influence. We also saw unique entries like WavLabLM, which were created from the ground up, independent of any pre-existing SSLs. The challenge's key findings include: (1) Scaling large models is not the only viable strategy for tackling multilingual speech tasks. (2) Diverse speech and voice types pose significant difficulties when applying multilingual speech representation to low-resource languages.




\section{Background}

\subsection{SUPERB and its Challenges}

Since the public release of SUPERB, researchers have widely adopted the benchmark, showcasing its increasing prominence. The speech SSL toolkit, S3PRL, designed to facilitate researchers' participation in SUPERB, has gained significant attention from researchers in self-supervised speech representation\footnote{\scriptsize{\url{https://github.com/s3prl/s3prl}}}. Reflecting the growing interest, the SUPERB team organized a dedicated SUPERB session during the 2nd Workshop on Self-supervised Learning for Audio and Speech Processing at AAAI 2022 \footnote{\scriptsize{\url{https://aaai-sas-2022.github.io/}}}. Additionally, some of the organizers conducted tutorials on SSL methodologies in speech and benchmarking with SUPERB at ICASSP 2022 and NAACL 2022 \footnote{\scriptsize{\url{https://sites.google.com/view/tutorial-ssl-speech}}}, further emphasizing the relevance and impact of SUPERB in the research community.

More recently, the SUPERB team successfully organized the SUPERB challenge at SLT 2023 \cite{feng2023superb}. The challenge received 12 speech SSL submissions, highlighting the continued interest and advancement in the field. Recognizing the importance of addressing efficiency concerns in speech SSL, the challenge introduced an additional examination of memory and computation estimation, supplementing the original SUPERB framework that primarily focused on the model's performance across different tasks. This expansion reflects the community's emphasis on optimizing the efficiency of speech SSL models alongside their task-specific capabilities.

\subsection{Multilingual Speech Self-supervised Representation}

Multilingual speech representation learning has received significant attention from both academia and industry. 

Before the advent of end-to-end ASR systems, numerous studies delved into multilingual representation for Hidden Markov Model (HMM)-based architectures \cite{6424246, vu2012multilingual, cui2015multilingual, sercu2017network}. With the evolution towards end-to-end ASR, researchers began investigating the use of large-scale multilingual end-to-end ASR models to learn generalized representations across multiple languages \cite{hou20_interspeech, pratap20c_interspeech, li2021scaling}. These studies underscored the importance of incorporating a broader range of languages and larger datasets to enhance performance, especially in low-resource ASR scenarios.

To further leverage the abundance of unlabeled data available in the wild, speech self-supervised models have been introduced in multilingual representation learning, offering the potential for even greater data utilization and performance gains. Kawakami et al., in 2020, leveraged various multilingual corpora with contrastive predictive coding (CPC), along with some English corpora, to jointly learn representations, resulting in significant performance enhancements across 22 languages \cite{kawakami2020learning}. More recently, Meta teams delved into multilingual representation learning using Transformer architecture, employing wav2vec~2.0, which led to notable advancements in the XLSR series of works \cite{conneau2020unsupervised, babu2021xls, duquenne2022speechmatrix}. These efforts exemplify the exploration and progress in leveraging self-supervised learning approaches for multilingual speech representation learning.

In recent works addressing low-resource or multilingual ASR, researchers have extensively explored the use of the XLSR model as a backbone, consistently achieving improvements over spectral features and other monolingual SSL approaches \cite{chen2023improving, yang2022jhu, arora2022espnet, tjandra2022improved, zhao2022improving, li2022asr2k, berrebbi22_interspeech}. However, there have been limited efforts to explore the integration of multilingual attributes into different self-supervised models, despite the success of WavLM \cite{chen2022wavlm} in outperforming wav2vec2 models across various speech processing tasks in SUPERB. This challenge, therefore, encourages the community to explore and develop improved architectures for multilingual speech processing.

In addition to ML-SUPERB, several other benchmarking initiatives have honed in on multilingual speech representation. LeBenchmark, for example, explores multilingual SSL in French speech processing \cite{evain21_interspeech}, while IndicSUPERB concentrates on a range of Indian languages \cite{javed2022indicsuperb}. Furthermore, XTREME-S covers an expansive array of tasks by directly amalgamating existing multilingual corpora without confining itself to unsupervised conditions \cite{conneau22_interspeech}. These benchmarking efforts provide invaluable resources and platforms for furthering research in multilingual speech representation.

Compared to other works, ML-SUPERB distinguishes itself by aiming to offer an efficient evaluation framework that covers an extensive range of languages and accounts for varying scenarios. To be precise, the public release of ML-SUPERB incorporates 143 languages, which, to the best of our knowledge, represents the benchmark with the most extensive language coverage. In terms of design, each language is equally sampled for ten minutes and one hour, thus curtailing the computational effort required for a complete evaluation cycle. With the introduction of the New Language Track, we anticipate an evolving,  ``open" ML-SUPERB that continues to integrate new languages.

\section{Tracks in ML-SUPERB Challenge}

In this challenge, participants are invited to engage in three different tracks. In addition to the research track, which welcomes regular research papers, the main challenge comprises two distinct tracks: the Challenge Track and the New Language Track. These tracks offer participants the opportunity to showcase their expertise and innovation in specific areas related to the challenge objectives.

\subsection{Challenge Track}

The Challenge Track serves as the primary focus of this challenge, aiming to explore new multilingual self-supervised models. Participants have the opportunity to compete on two leaderboards: a public leaderboard and a hidden leaderboard.

The public leaderboard utilizes the ML-SUPERB benchmark data, which was released in \cite{shi2023ml}. On the other hand, the hidden leaderboard incorporates data submitted from the New Language Track. The evaluation methodology of ML-SUPERB follows a similar design to that of SUPERB. It utilizes a frozen upstream SSL model and a fixed downstream model architecture. Specifically, a weighted sum of layer-wise SSL features is employed, and a connectionist-temporal classification (CTC)-based transformer network is used as the downstream model.\footnote{Implementation details are at \scriptsize{\url{https://github.com/espnet/espnet/tree/master/egs2/ml_superb/asr1}}.}

Both leaderboards consist of four tasks: monolingual ASR, multilingual ASR, LID, and multilingual ASR+LID. Each task has two configurations, with a 10-minute training set and a 1-hour training set, respectively. Additionally, the public leaderboard includes few-shot learning cases, where only 5 utterances from 20 languages are used in the training set. However, the hidden set does not consider this specific case.

\subsection{New Language Track}

In addition to the Challenge Track, the ML-SUPERB Challenge presents a unique New Language Track. This track is specifically tailored for researchers focused on language resources, particularly those keen on evaluating their low-resource language data employing cutting-edge ASR techniques. The principal aim of this track is to encourage researchers to contribute their unique language data to ML-SUPERB, consequently broadening the spectrum of multilingual research to encapsulate a wider variety of global languages. Participants engaging in the New Language Track are obligated to offer a comprehensive description of their submitted data and to execute experiments utilizing established models from the ML-SUPERB benchmark.

\begin{figure}[tbp]
    \centering
    \includegraphics[width=\linewidth]{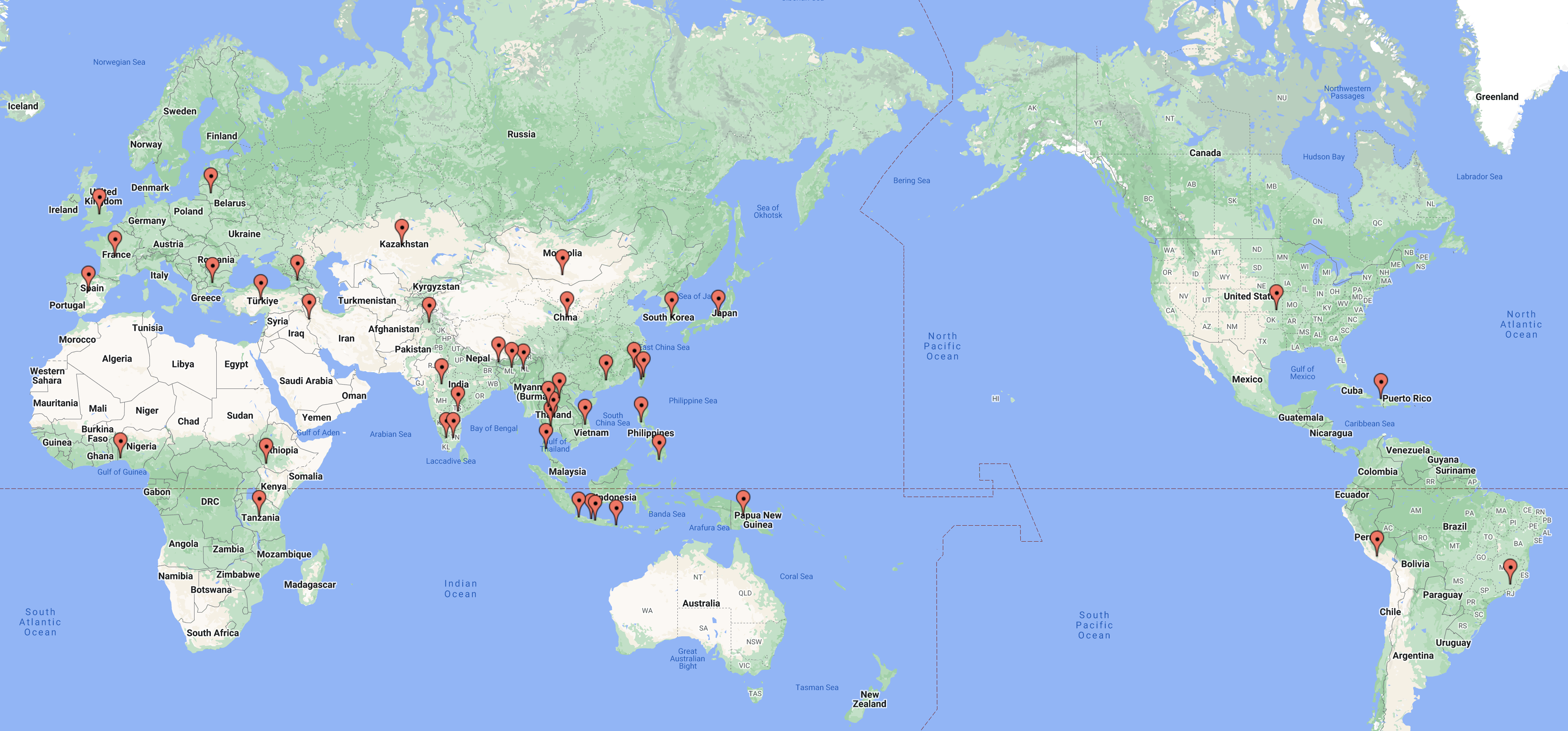}
    \vspace{-15pt}
    \caption{Geographical distribution of the New Language track submissions. The 45 languages are marked on a map with their rough locations of speaking.}
    \label{fig: map-new-lang}
    \vspace{-15pt}
\end{figure}

An important aspect of the New Language Track is that the submitted data serves as an open evaluation set for other participants who have submitted their own SSL models to the challenge. This establishes collaboration and allows participants to evaluate their models on a diverse set of languages, contributing to a more comprehensive and inclusive evaluation of multilingual speech recognition.

\section{Submissions}

\subsection{New Language Track Submissions}

For the New Language track, the challenge received a total of 8 submissions. In combination with the base evaluation hidden set prepared by the organizers, these submissions resulted in the creation of 54 new additional ML-SUPERB-style corpora. Figure~\ref{fig: map-new-lang} illustrates the distribution of these hidden languages, which are primarily concentrated in East Asia \cite{yihuiasru}, Southeast Asia \cite{saktiasru}, and South Asia, while also exhibiting a reasonable distribution across other regions of the world (\`{I}r\`{o}y\`{i}nSpeech \cite{ogunremi2023r}, Quechua Speech \cite{chen2023evaluating}, etc.). A number of submissions are created with existing corpora \cite{thaidialect_interspeech2023, sakti-2013, cahyawijaya-etal-2023-nusacrowd, sakti-2013, cardenas2018siminchik}, while there are also effects in releasing newly published low-resource languages to the challenge.


In addition to the submissions received for the New Language Track, the organizers have prepared additional data for evaluation purposes, known as the base-hidden set. The base-hidden set consists of two main concentrations: multilingual conversational speech and multilingual singing voice. These additions create more challenging scenarios for multilingual understanding and evaluation. The conversational speech samples are drawn from various sources, including Babel \cite{roach1996babel}, Fisher \cite{cieri2004fisher}, Switchboard \cite{godfrey1992switchboard}, KsponSpeech \cite{bang2020ksponspeech}, and AccentedFrench\footnote{\scriptsize{\url{https://www.openslr.org/57/}}}. On the other hand, the singing voice samples are taken from Muskits recipes \cite{shi2022muskits}, including Opencpop \cite{wang22b_interspeech}, PopCS \cite{huang2021multi}, M4Singer \cite{zhang2022m4singer}, CSD \cite{choi2020children}, as well as a combination of the Oniku and Ofuton corpora\footnote{Gotanmiya Kurumi Singing Voice Database: \scriptsize{\url{http://onikuru.info/db-download/}} and Ofuton P Singing Voice Database: \scriptsize{\url{https://sites.google.com/view/oftn-utagoedb}}}. The inclusion of these additional datasets in the base-hidden set enhances the complexity and diversity of the evaluation, providing more realistic and challenging scenarios for evaluating multilingual speech understanding systems.

\begin{table}[t]
    \centering
    \caption{Benchmark statistics on the hidden leaderboard from the New Language track.}
    \include{tables/dataset}
    \label{tab: dataset}
    \vspace{-15pt}
\end{table}

The statistics of the hidden set are detailed in Table~\ref{tab: dataset}. Out of the 54 newly incorporated ML-SUPERB-style corpora, a total of 18 have been deemed fitting for integration into the future public benchmark of ML-SUPERB. These 18 corpora will serve to broaden and enrich the publicly available benchmark. The remaining corpora have been earmarked for internal evaluation within the ML-SUPERB framework. With this newly established schema, we anticipate that ML-SUPERB will continue to evolve by perpetually adding new languages, fostering a better ecosystem for multilingual speech research worldwide.

\begin{table}[t]
    \centering
    \caption{Selected models from the public ML-SUPERB and challenge submissions from participants. Different colors represent different pre-trained languages: purple stands for monolingual SSL, blue stands for SSL trained in a few languages from the same region, and yellow stands for multilingual SSLs. }
    \include{tables/model}
    \label{tab: models}
    \vspace{-15pt}
\end{table}

\subsection{Challenge Submissions}

\begin{table*}[!ht]
    \centering
    \caption{\{10-minute / 1-hour\} set ML-SUPERB public benchmark (143 languages).}
    \include{tables/10min-public-set}
    \label{tab: 10min}
    \vspace{-15pt}
\end{table*}


\begin{table*}[!ht]
    \centering
    \caption{\{10-minute / 1-hour\} set ML-SUPERB hidden benchmark (54 languages).}
    \include{tables/10min-hidden-set}
    \label{tab: 10min-hidden}
    \vspace{-15pt}
\end{table*}


For the ML-SUPERB challenge, we received 12 model submissions, shown in Table~\ref{tab: models}. For readers' reference, the organizers also present the results of six example models in the original public benchmark.\footnote{The six models are selected based on their relative performances over the existing public benchmark reported in \cite{shi2023ml}.} The followings are their brief descriptions, which are categorized by their pre-training methods:

\noindent \textbf{HuBERT with multiple resolutions}: The set of models takes the insight from \cite{shi2023exploration} by utilizing HuBERT with multiple resolutions \cite{chen2023joint}. In its pre-training stage, the participants utilized CommonVoice 11.0 dataset \cite{ardila2020common} and extracted the K-means units with the English HuBERT released in \cite{chen2023reducing}. Similar to \cite{shi2023exploration}, the participants trained three HuBERT-based on 20ms, 40ms, and 80ms resolution by controlling the convolutional feature extractor. To form a HuBERT with Multiple Resolution (HuBERT-MR), the participants further combined the three pre-trained HuBERT by simply concatenating them with upsampling. Following \cite{shi2023exploration}, the upsampling is simply repeating without additional introduction of learnable parameters. Four submissions are received from the participants, including three CommonVoice-HuBERT (i.e., CV-HuBERT)-base models trained on 20ms, 40ms, and 80ms resolutions and a CV-HuBERT-MR of the combination of all three HuBERT with different resolutions.

\noindent \textbf{Ensemble modeling}: The ensemble modeling is straightforward by stacking the representation of two SSL models. The implementation concept is related to the previous investigation in SSL model fusion \cite{berrebbi22_interspeech, chen22q_interspeech}. The method does not introduce additional parameters as previous works \cite{berrebbi22_interspeech, chen22q_interspeech}, but instead has the requirement of the same representation resolution across different SSL models. The EFFUSE team submitted two models, including EFFUSE (wav2vec2+XLSR) and EFFUSE (HuBERT+XLSR)  \cite{effuse}.

\begin{figure*}
\centering
\begin{tikzpicture}
\begin{axis}[
    xlabel={MACs (G)},
    ylabel={SUPERB$_{s}$},
    width=12cm,
    height=6cm,
    scaled x ticks=false,
    legend style={at={(0.5,-0.25)},anchor=north, legend columns=5, font=\footnotesize},
        legend cell align={left},
    grid=both,
    xmin=500, xmax=15000, ymin=150, ymax=1000
]
\addplot[
    scatter,
    only marks,
    scatter src=explicit symbolic,
    scatter/classes={
        wav2vec2-base={mark=square*,color1},
        wav2vec2-base-23={mark=square*,color2},
        XLSR-128={mark=square*,color3},
        HuBERT-base={mark=square*,color4},
        HuBERT-large={mark=square*,color5},
        mHuBERT-base={mark=square*,color6},
        MMS-300m={mark=square*,color7},
        MMS-1b={mark=square*,color8},
        CV-HuBERT-base={mark=square*,color9},
        CV-HuBERT-base(40ms)={mark=square*,color10},
        CV-HuBERT-base(80ms)={mark=square*,color11},
        CV-HuBERT-MR-base={mark=square*,color12},
        EFFUSE(W2V2+XLSR)={mark=square*,color13},
        EFFUSE(HuBERT+XLSR)={mark=square*,color14},
        NWHC1={mark=square*,color15},
        NWHC2={mark=square*,color16},
        WavLabLM-base={mark=square*,color17},
        WavLabLM-large-EK={mark=square*,color18},
        WavLabLM-large-MK={mark=square*,color19},
        WavLabLM-large-MS={mark=square*,color20}
    },
] table [meta=label,x=x,y=y] {mac-superb.txt};
\node (label1) at (axis cs:1669,620.3) [pin=280:1] {};
\node (label2) at (axis cs:1669,654.2) [pin=300:2] {};
\node (label3) at (axis cs:4328,962.0) [pin=340:3] {};
\node (label4) at (axis cs:1669,760.5) [pin=300:4] {};
\node (label5) at (axis cs:4324,730.9) [pin=330:5] {};
\node (label6) at (axis cs:1669,771.2) [pin=320:6] {};
\node (label7) at (axis cs:4328,774.7) [pin=340:7] {};
\node (label8) at (axis cs:12014,933.9) [pin=300:8] {};
\node (label9) at (axis cs:1666,857.3) [pin=340:9] {};
\node (label10) at (axis cs:904,547.6) [pin=300:10] {};
\node (label11) at (axis cs:587,179.0) [pin=30:11] {};
\node (label12) at (axis cs:3156,804.8) [pin=300:12] {};
\node (label13) at (axis cs:8648,826.9) [pin=300:13] {};
\node (label14) at (axis cs:8648,865.0) [pin=330:14] {};
\node (label15) at (axis cs:4328,966.5) [pin=360:15] {};
\node (label16) at (axis cs:4328,961.8) [pin=300:16] {};  
\node (label17) at (axis cs:1666,772.1) [pin=350:17] {};
\node (label18) at (axis cs:4328,567.9) [pin=300:18] {};
\node (label19) at (axis cs:4328,672.4) [pin=330:19] {};
\node (label20) at (axis cs:4328,659.7) [pin=300:20] {};
\legend{1. wav2vec2-base, 2. wav2vec2-base-23, 3. XLSR-128, 4. HuBERT-base, 5. HuBERT-large, 6. mHuBERT-base, 7. MMS-300m, 8. MMS-1b, 9. CV-HuBERT-base, 10. CV-HuBERT-base(40ms), 11. CV-HUBERT-base(80ms), 12. CV-HUBERT-MR-base, 13. EFFUSE(W2V2+XLSR), 14. EFFUSE(HuBERT+XLSR), 15. NWHC1, 16. NWHC2, 17. WavLabLM-base, 18. WavLabLM-large-EK, 19. WavLabLM-large-MK, 20. WavLabLM-large-MS}
\end{axis}
\end{tikzpicture}
\caption{\label{fig: macs}MACs v.s. SUPERB score in ML-SUPERB 1-hour hidden benchmark.}
\end{figure*}

\noindent \textbf{Parameter-level modification}: The NWHC team's submissions introduce an inventive strategy aimed at preserving a higher volume of content information derived from SSL models \cite{xue2023SSHR}. Their hierarchical representation analysis, executed during ASR training, revealed a continual decrease in content information. This was particularly apparent in the diminished ASR performance within the concluding layers. To enhance the performance of downstream tasks involving the SSL, they propose a modification to the Massively Multilingual Speech (MMS)-300m model. Specifically, they replace the last few layers with intermediate layers: for NWHC1, the final three layers are supplanted by the 17th-19th layers, while for NWHC2, the last four layers are supplanted by the 17th-20th layers. Crucially, this modification is implemented directly at the level of network parameters, rather than at the resultant representation.

\noindent \textbf{WavLM-style pre-training}: WavLabLM are submitted by the WAVLab team \cite{chen2023joint}, by adopting WavLM-style pre-training \cite{chen2022wavlm} into multilingual scenarios. A noisy speech simulation protocol is applied to the pre-training by mixing utterances and noises. The pre-training data combined several open-source corpora, reaching around 40k hours of speech over 136 languages. Four models are submitted for the challenge, including a base model with less parameter size and three large models with different training targets and strategies. All four models are trained with the k-means clusters from their previous HuBERT-large model \cite{chen2023reducing}. WavLabLM-large-EK utilizes the English data to extract k-means targets, while the other two models utilize multilingual data to extract targets. WavLabLM-large-MS further adopts a multi-stage training strategy by upsampling low-resource languages within the dataset. Note that WavLabLMs are distinctive submissions that produce pre-training SSL models from the ground up, without relying on pre-existing SSLs. These models were developed using academic computing resources, and their source codes have been made publicly available to researchers. This makes it easier for individuals who do not have access to advanced computing resources to conduct pre-training research on SSL.

In addition to the submissions, the organizers also include the evaluation of pre-trained wav2vec2.0 presented in the Massively Multilingual Speech (MMS) project from Meta \cite{pratap2023scaling}. Therefore, in total, the evaluation of this ML-SUPERB challenge adds up to 14 new models to the benchmark.

\section{Challenge Results Summary}

The public leaderboard results are presented in Table~\ref{tab: 10min}, while the hidden leaderboard results are shown in Table~\ref{tab: 10min-hidden}.

\noindent \textbf{Overall}: Across both the 10-minute and 1-hour public benchmarks (Table~\ref{tab: 10min}), MMS-1b delivered the best SUPERB score, significantly outpacing its competitors. Given the wide coverage of languages, it is unsurprising to observe the strong generalization capability of the MMS-based model in multilingual speech tasks, particularly in few-shot tasks. In the hidden benchmarks (Table~\ref{tab: 10min-hidden}), MMS-1b excels in the 10-minute benchmark, but was overtaken by the XLSR-128 and NWHC models in the 1-hour benchmark, especially in the LID and multilingual ASR+LID tasks. Remarkably, NWHC2 achieves the best performance in the 1-hour hidden leaderboard, outperforming XLSR-128 and NWHC1.

\noindent \textbf{HuBERT with multiple resolutions}: In all scenarios, CV-HuBERT-base utilizing the default 20ms resolution units outperformed the original HuBERT-base model, emphasizing the importance of multilingual pre-training. Contrary to the observation in \cite{shi2023exploration}, HuBERT with multiple resolutions did not invariably enhance performance, but often undermined the performance of the CV-HuBERT-base. This suggests that techniques used in monolingual contexts may not always translate effectively to multilingual scenarios due to linguistic variation and phonetic distribution differences.

\noindent \textbf{Ensemble modeling}: Despite being a prevalent strategy in numerous challenges, naive ensemble modeling by concatenating SSL features does not always improve multilingual task performance. Specifically, in this challenge, ensemble modeling's performance seems largely limited by weaker SSL models. Consequently, both ensemble models (i.e., EFFUSE models) did not significantly outperform XLSR-128, even though the XLSR-128 representation is included in the ensemble frameworks.

\noindent \textbf{Parameter-level modification}: Despite being based on modified versions of MMS-300m, the NWHC systems surpassed MMS-300m across all benchmarks, even achieving the best performance in the 1-hour hidden benchmark. These impressive results underscore the significance of layer-wise analysis of self-supervised models and raise questions about how to fully leverage SSL models.

\noindent \textbf{WavLM-style pre-training}: While WavLM has been a top performer in the SUPERB benchmark \cite{yang21c_interspeech}, one may wonder whether similar pre-training approaches could excel in multilingual scenarios. The analysis of WavLabLMs revealed that the impact of denoising modeling in ML-SUPERB may not be as beneficial as in SUPERB. While the model series outperformed the original HuBERT-based model in the public benchmark, their performance deteriorated in the hidden benchmarks, especially in multilingual tasks.

\noindent \textbf{Public benchmarks vs. hidden benchmarks}: As per \cite{shi2023ml}, the majority of the public benchmark comprises read speech, while the hidden benchmark includes a substantial set of conversational speech and singing voices. These varied voice styles present significant challenges for the ML-SUPERB tasks. The performance in the hidden sets is considerably worse than that in the public sets. Despite generally similar rankings, some SSL model performances vary. For instance, MMS-1b, despite being the top performer in all other leaderboards, did not achieve the best performance in the 1-hour hidden benchmark. As in the real world, researchers focusing on low-resource languages often have limited access to clean-read speech. Therefore, studying multilingual representation across different voice types could become a major research direction.

\noindent \textbf{Multilingual vs. monolingual}: In the inaugural ML-SUPERB release, most models were either monolingual-based or focused on a small set of languages, with limited exploration of multilingual SSL models. For the 2023 ML-SUPERB challenge, all model submissions underwent pre-training in multilingual data spanning over 50 languages. The leaderboard results clearly indicate that multilingual SSL typically outperforms those trained with a limited language scope, suggesting that focusing on multilingual representation for multilingual tasks is a promising future research direction.

\noindent \textbf{Performance vs. efficiency}: Taking a cue from the SUPERB challenge at SLT2022 \cite{feng2023superb}, we've also considered efficiency as a critical factor when evaluating self-supervised models in the ML-SUPERB challenge. We've gauged the theoretical multiply-accumulate operation (MACs) based on the profiling toolkit utilized in the prior SUPERB challenge\footnote{\scriptsize{\url{https://github.com/B06901052/DeepSpeed/tree/superb-challenge}}}. The tradeoff between model computational complexity (measured by MACs) and ML-SUPERB performance (indicated by the SUPERB score) is illustrated in Fig~\ref{fig: macs}.

In the previous SUPERB challenge \cite{feng2023superb}, we observed a scaling rule where increasing model size typically led to improved performances across various speech processing tasks. However, it is evident that the scaling rule isn't always effective for multilingual tasks. Specifically, smaller models can sometimes deliver performance equal to or better than their larger counterparts. For instance, CV-HuBERT-base outperforms MMS-300m and HuBERT-large, even though it utilizes a base architecture. Similarly, NWHC2 slightly outperforms MMS-1b in the 1-hour hidden benchmark, despite its smaller size and computational burden. In more comparable scenarios, WavLabLM-base outperforms all three WavLabLM-large models, with model size being the only differentiating factor. These observations hint that large-scale SSL pre-training might not be the only viable path for multilingual SSL representation learning.

\section{Conclusion}

The ML-SUPERB challenge of 2023 has provided a platform for exploring and developing multilingual speech SSL models in multilingual ASR and LID. The challenge attracted wide-ranging participation, yielding valuable insights into the state-of-the-art methods and potential future directions for this emerging field.

Findings from the challenge underscore that while large model scaling can be effective, it is not the exclusive solution to advancing multilingual speech tasks. It was observed that smaller models could potentially deliver comparable or even superior performance, highlighting the potential for efficient and effective model development.  The challenge also uncovered the major difficulties in tackling varying speech and voice types, especially in low-resource languages, pointing to a crucial research direction for future multilingual representation learning endeavors.

The results of this year's ML-SUPERB challenge also reinforce the notion that multilingual SSL models usually outperform those trained with limited language coverage. Therefore, the path towards multilingual representation for multilingual tasks stands out as a promising direction to further explore.

As the field continues to evolve, the insights gathered from this challenge will serve as valuable stepping stones, informing and directing future research efforts in multilingual SSL model development. We expect the ML-SUPERB challenge to continue to play a pivotal role in shaping this fascinating and crucial area of research.

\section{Acknowledgements}
Some experiments of this work used the Bridges2 system at PSC and Delta system at NCSA through allocation CIS210014 from the Advanced Cyberinfrastructure Coordination Ecosystem: Services \& Support (ACCESS) program, which is supported by National Science Foundation grants \#2138259, \#2138286, \#2138307, \#2137603, and \#2138296. We also gratefully acknowledge the support of NVIDIA Corporation with the donation of the A6000 GPUs used for this research.

\bibliographystyle{IEEEbib}
\bibliography{strings,refs}

\end{document}

%% file: tables/dataset.tex




    \begin{tabular}{l|c|c}
        \toprule
Dataset & Hours & Normal Langs (45) \\
\midrule
 10-minute & 9.52 & $\sim$10min $\times$ 54 (\texttt{lang}, \texttt{data})  \\
 1-hour & 57.13 & $\sim$1h $\times$ 54 (\texttt{lang}, \texttt{data}) \\
 Dev  & 9.31 & $\sim$10min $\times$ 54 (\texttt{lang}, \texttt{data})  \\
 Test  & 9.38 & $\sim$10min $\times$ 54 (\texttt{lang}, \texttt{data})  \\
        \bottomrule
    \end{tabular}


%% file: tables/model.tex
\resizebox {0.8\linewidth} {!} {

    \begin{tabular}{l|c|cc}
        \toprule
        \multirow{2}{*}{Model} & \multirow{2}{*}{Params (M)} &  \multicolumn{2}{c}{Pre-Training}   \\
         &  &  \# Hours &  \# Langs \\
\midrule
\rowcolor{mono} wav2vec2-base \cite{baevski2020wav2vec}  & 95 & 1k & 1 \\
\rowcolor{regional} wav2vec2-base-23 \cite{wang2021voxpopuli}  & 95 & 100k & 23 \\
\rowcolor{multi} XLSR-128 \cite{babu2021xls}  & 317  & 400k & 128 \\
\rowcolor{mono} HuBERT-base \cite{hsu2021hubert} & 95 & 1k & 1  \\
\rowcolor{mono} HuBERT-large \cite{hsu2021hubert}  & 317 & 60k & 1 \\
\rowcolor{regional} mHuBERT-base \cite{lee2022textless}  & 95 & 14k & 3 \\
\midrule
\rowcolor{multi} MMS-300m & 317 & 491k & 1,406 \\
\rowcolor{multi} MMS-1b & 965 & 491k & 1,406 \\
\midrule
\rowcolor{multi} CV-HuBERT-base & 95 & 13k & 92 \\
\rowcolor{multi} CV-HuBERT-base (40ms) & 96 & 13k & 92  \\
\rowcolor{multi} CV-HuBERT-base (80ms) & 96 & 13k & 92 \\
\rowcolor{multi} CV-HuBERT-MR-base & 287 & 13k & 92 \\
\midrule
\rowcolor{multi} EFFUSE (W2V2+XLSR) & 634 & 400k & 128 \\
\rowcolor{multi} EFFUSE (HuBERT+XLSR) & 634 & 400k & 128 \\
\midrule
\rowcolor{multi} NWHC1 & 317 & 400k & 128 \\
\rowcolor{multi} NWHC2 & 317 & 400k & 128 \\
\midrule
\rowcolor{multi} WavLabLM-base & 95 & 40k & 136\\
\rowcolor{multi} WavLabLM-large-EK & 317 & 40k & 136  \\
\rowcolor{multi} WavLabLM-large-MK & 317 & 40k & 136 \\
\rowcolor{multi} WavLabLM-large-MS & 317 & 40k & 136 \\
        \bottomrule
    \end{tabular}

}

%% file: tables/10min-public-set.tex
\resizebox {\linewidth} {!} {
\begin{tabular}{l|c|cc|c|ccc|c}
\toprule
\multirow{3}{*}{SSL} & Monolingual ASR & \multicolumn{2}{c|}{Multilingual ASR} & \multicolumn{1}{c|}{LID} & \multicolumn{3}{c|}{Multilingual ASR + LID} & \multirow{3}{*}{SUPERB$_{s}$} \\
&          &             Normal & Few-shot & Normal & \multicolumn{2}{c}{Normal} & \multicolumn{1}{c|}{Few-shot} \\
& CER/PER & CER & CER & ACC & ACC & CER & \multicolumn{1}{c|}{CER} &  \\
\midrule
FBANK & 72.1 / 63.7 & 62.4 / 59.3 & 58.3 / 57.4 & 11.1 / 9.3 & 35.9 / 43.5 & 62.0 / 58.6 & 58.9 / 58.1 & 0 / 0 \\
\rowcolor{mono} wav2vec2-base \cite{baevski2020wav2vec} & 44.2 / 35.9 & 43.0 / 35.5 & 45.7 / 44.3 & 54.4 / 80.8 & 66.9 / 83.6 & 40.6 / 32.1 & 44.2 / 42.6 & 590.4 / 707.8 \\
\rowcolor{regional} wav2vec2-base-23 \cite{wang2021voxpopuli} & 49.2 / 35.1  & 37.7 / 32.0 & 43.4 / 42.2 & 58.7 / 71.9 & 45.1 / 66.3 & 37.2 / 30.9 & 44.3 / 43.0 & 563.0 / 676.5 \\
\rowcolor{multi} XLSR-128 \cite{babu2021xls} & 39.7 / 30.6 & 29.2 / 22.0 & 40.9 / 39.3 & 66.9 / 87.9 & 55.6 / 85.6 & 28.4 / 22.9 & 42.1 / 42.4 & 734.9 / 854.2 \\
\rowcolor{mono} HuBERT-base \cite{hsu2021hubert} & 42.8 / 35.3 & 39.8 / 31.4 & 44.5 / 42.7 & 61.2 / 86.1 & 71.5 / 86.0 &  39.2 / 30.9 & 43.8 / 41.8 & 650.8 / 757.0 \\
\rowcolor{mono} HuBERT-large \cite{hsu2021hubert} & 38.2 / 32.2 & 44.4 / 37.7 & 48.2 / 43.5 &  46.5 / 64.1 & 55.4 / 77.7 & 45.6 / 35.1 & 49.3 / 42.2 & 541.8 / 659.8 \\
\rowcolor{regional} mHuBERT-base \cite{lee2022textless} & 41.0 / 33.0 & 40.5 / 33.4 & 45.6 / 43.6 & 52.4 / 72.5  & 46.6 / 70.9 & 36.8 / 29.7 & 44.2 / 43.1 & 580.3 / 692.1 \\
\midrule
\rowcolor{multi} MMS-300m & 33.8 / 30.5 & 28.7 / 24.0 & 36.5 / 36.5 & 62.3 / 84.3 & 71.9 / 74.3 & 31.5 / 30.0 & 30.9 / 29.2 & 826.7 / 841.3 \\
\rowcolor{multi} MMS-1b & \textbf{33.3} / \textbf{25.7} & \textbf{21.3} / \textbf{18.1} & \textbf{30.2} / \textbf{30.8} & \textbf{84.8} / 86.1 & 73.3 / 74.8 & \textbf{26.0} / 25.5  & \textbf{25.4} / \textbf{24.8} & \textbf{983.5} / \textbf{943.2} \\
\midrule
\rowcolor{multi} CV-HuBERT-base & 41.9 / 32.9 & 35.4 / 27.5 & 44.0 / 40.8 & 71.2 / 84.0 & 76.6 / 87.3 & 35.1 / 28.2 & 43.6/ 41.1 & 726.3 / 796.7 \\
\rowcolor{multi} CV-HuBERT-base (40ms) & 71.6 / 62.6 & 60.5 / 52.0 & 57.5 / 53.0 & 65.6 / 83.0 & 65.7 / 83.3 & 59.6 / 52.3 & 57.7 / 53.4 & 179.6 / 380.4 \\
\rowcolor{multi} CV-HuBERT-base (80ms) & 76.4 / 67.6 & 72.7 / 70.7 & 66.1 / 64.1 & 33.2 / 57.2 & 17.2 / 39.4 & 72.3 / 70.4 & 64.2 / 64.4 & 130.4 / 16.2 \\
\rowcolor{multi} CV-HuBERT-MR-base & 47.8 / 38.3 & 37.0 / 28.3 & 43.2 / 40.8 & 64.1 / 86.0 & 74.8 / 84.5 & 36.2 / 30.6 & 42.5 / 41.0 & 659.0 / 755.0 \\
\midrule
\rowcolor{multi} EFFUSE (W2V2+XLSR) & 38.5 / 28.9 & 31.0 / 24.4 & 40.9 / 40.2 & 23.0 / 13.5 &  69.7 / 87.8 & 31.4 / 24.4 & 41.8 / 39.3 & 610.2 / 624.6 \\
\rowcolor{multi} EFFUSE (HuBERT+XLSR) & 37.9 / 29.5 & 31.8 / 23.5 & 42.5 / 38.3 & 59.4 / 79.0 & 72.7 / 89.9 & 31.2 / 23.3 & 41.0 / 37.6 & 736.1 / 849.8  \\
\midrule
\rowcolor{multi} NWHC1 & 39.5 / 30.5 & 28.9 / 21.5 & 41.4 / 38.6 & 67.1 / 87.4 & 77.1 / \textbf{90.6} & 28.8 / \textbf{21.5} & 40.3 / 38.2 & 781.5 / 878.6 \\
\rowcolor{multi} NWHC2 & 39.5 / 30.5 & 29.3 / 21.6 & 42.0 / 39.3 & 64.4 / \textbf{88.1} & \textbf{77.4} / \textbf{90.6} & 28.4 / 21.8 & 41.5 / 38.8 & 767.5 / 875.3  \\
\midrule
\rowcolor{multi} WavLabLM-base & 45.6 / 37.6 & 45.3 / 39.6 & 45.7 / 44.5 & 40.7 / 56.5 & 51.8 / 67.8 & 44.3 / 38.2 & 44.7 / 43.9 & 488.0 / 561.8 \\
\rowcolor{multi} WavLabLM-large-EK & 40.7 / 33.7 & 41.0 . 33,5 & 44.1 / 41.9 & 61.2 / 83.4 & 60.0 / 79.9 & 40.0 / 33.1 & 42.6 / 41.3 & 640.0 / 741.7 \\
\rowcolor{multi} WavLabLM-large-MK & 40.5 / 32.3 & 38.8 / 32.8 & 44.4 / 42.8 & 67.6 / 79.0 & 69.0 / 79.6 & 38.6 / 32.8 & 44.2 / 42.4 & 686.4 / 732.7\\
\rowcolor{multi} WavLabLM-large-MS & 40.5 / 32.8 & 37.8 / 31.9 & 43.8 / 42.8 & 71.7 / 81.1 & 70.8 / 80.0 & 37.0 / 32.2 & 43.4 / 41.2 & 715.0 / 743.3 \\
\bottomrule
\end{tabular}
}

%% file: tables/10min-hidden-set.tex
\resizebox {0.9\linewidth} {!} {
\begin{tabular}{l|c|c|c|cc|c}
\toprule
\multirow{2}{*}{SSL} & Monolingual ASR & Multilingual ASR & LID & \multicolumn{2}{c|}{Multilingual ASR + LID} & \multirow{3}{*}{SUPERB$_{s}$} \\
& CER/PER & CER & ACC & ACC & CER &   \\
\midrule
FBANK & 76.4 / 70.7 & 71.9 / 68.4 & 21.8 / 14.7 & 29.5 / 37.1 & 70.4 / 65.8 & 0 / 0\\
\rowcolor{mono} wav2vec2-base \cite{baevski2020wav2vec} & 62.7 / 54.2 & 55.0 / 47.0 & 43.0 / 60.2 & 47.0 / 42.2 & 54.3 / 42.6 & 585.3 / 620.3  \\
\rowcolor{regional} wav2vec2-base-23 \cite{wang2021voxpopuli} & 63.7 / 56.1 & 55.4 / 49.9 & 44.1 / 57.0 & 48.8 / 59.9 & 55.0 / 48.5 & 581.8 / 654.2 \\
\rowcolor{multi} XLSR-128 \cite{babu2021xls} & 57.6 / 49.2 & 47.9 / 39.0 & 48.4 / \textbf{70.2} & 51.2 / \textbf{70.7}  & 47.0 / \textbf{37.7} & 789.4 / 962.0\\
\rowcolor{mono} HuBERT-base \cite{hsu2021hubert} & 61.1 / 53.8 & 54.9 / 47.2 & 47.5 / 65.1 & 46.8 / 63.1 & 53.6 / 45.8 & 637.2 / 760.5\\
\rowcolor{mono} HuBERT-large \cite{hsu2021hubert} & 62.7 / 52.3 & 54.3 / 47.3 & 44.4 / 57.9 & 43.2 / 58.9 & 53.3 / 45.0  & 589.1 / 730.9  \\
\rowcolor{regional} mHuBERT-base \cite{lee2022textless} & 59.8 / 53.3 & 53.2 / 46.1 & 45.2 / 65.3 & 44.0 / 61.6 & 52.7 / 45.5 & 644.0 / 771.2 \\
\midrule
\rowcolor{multi} MMS-300m & 60.1 / 51.0 & 47.3 / 42.2 & 48.9 / 45.3 & 55.4/ 66.2 & 46.2 / 40.8 & 788.6 / 774.7 \\
\rowcolor{multi} MMS-1b & \textbf{55.4} / \textbf{46.8} & \textbf{42.0} / \textbf{37.4} & \textbf{59.4} / 65.4 & \textbf{60.0} / 60.9 & \textbf{40.9} / 39.5 & \textbf{1000.0} / 933.9 \\
\midrule
\rowcolor{multi} CV-HuBERT-base & 59.1 / 52.3 & 52.0 / 43.7 & 48.9 / 68.9 & 51.0 / 68.9 & 50.9 / 42.6 & 723.3 / 857.3 \\
\rowcolor{multi} CV-HuBERT-base (40ms) & 71.0 / 63.2 & 64.9 / 52.0 & 47.3 / 66.0 & 40.4 / 57.4 & 64.5 / 59.0 & 362.0 / 547.6  \\
\rowcolor{multi} CV-HuBERT-base (80ms) & 72.5 / 67.9 & 76.4 / 69.8 & 28.3 / 50.9 & 28.3 / 40.5 & 71.0 / 69.1 & 44.6 / 179.0\\
\rowcolor{multi} CV-HuBERT-MR-base & 62.6 / 54.8 & 54.3 / 44.5 & 47.3 / 55.9 & 47.3 / 66.4 & 53.0 / 39.3 & 627.7 / 804.8  \\
\midrule
\rowcolor{multi} EFFUSE (W2V2+XLSR) & 61.3 / 50.6 & 49.6 / 41.9 & 45.0 / 57.3 & 49.5 / 63.7 & 48.8 / 40.6 & 694.0 / 826.9 \\
\rowcolor{multi} EFFUSE (HuBERT+XLSR) & 57.5 / 49.1 & 50.7 / 40.3 & 48.2 / 55.9 & 48.9 / 66.4 & 49.2 / 39.3 & 747.1 / 865.0 \\
\midrule
\rowcolor{multi} NWHC1 & 56.7 / 47.7 & 48.2 / 38.9 & 47.6 / 69.4 & 51.5 / 69.4 & 47.0 / 38.5 & 793.5 / \textbf{966.5} \\
\rowcolor{multi} NWHC2 & 56.6 / 47.8 & 47.9 / 38.9 & 47.6 / 68.9 & 51.5 / 68.9 & 47.1 / 38.4 & 796.8 / 961.8 \\
\midrule
\rowcolor{multi} WavLabLM-base & 55.9 / 47.2 & 56.6 / 47.6 & 45.4 / 58.2 & 45.7 / 58.1 & 54.7 / 46.8 & 661.8 /772.1 \\
\rowcolor{multi} WavLabLM-large-EK & 57.2 / 50.3 & 62.9 / 54.2 & 46.9 / 55.7 & 45.2 / 45.4 & 60.3 / 60.3 & 577.9 / 567.9 \\
\rowcolor{multi} WavLabLM-large-MK & 56.9 / 50.3 & 64.1 / 55.6 & 49.3 / 62.0 & 49.3 / 61.0 & 61.6 / 53.7 & 598.6 / 672.4 \\
\rowcolor{multi} WavLabLM-large-MS & 59.3 / 50.7 & 61.6 / 55.5 & 49.7 / 61.6 & 52.3 / 59.2 & 58.9 / 53.9 & 617.4 / 659.7 \\

\bottomrule
\end{tabular}
}